\documentclass[prl,twocolumn,superscriptaddress,nofootinbib]{revtex4}

\usepackage{graphicx,bm,times}
\usepackage{tabularx} 
\usepackage{lipsum}
\usepackage{amsmath}
\usepackage{gensymb}
\usepackage{color,soul, mathrsfs}
\usepackage[]{natbib}
\usepackage{mathtools}
\usepackage[pdftex,colorlinks=true,linkcolor=blue,citecolor=blue,urlcolor=black]{hyperref}
\usepackage{amsmath, amsthm, amssymb}
\usepackage{subfigure}
\usepackage{comment}
\usepackage{url}
\usepackage{tikz}

\usepackage{gensymb}

\usepackage{bbold}

\newcommand{\ket}[1]{| #1 \rangle}

\newcommand{\myfont}[1]{\fontfamily{AnonymousPro}\selectfont}
\DeclareTextFontCommand{\myfontcommand}{\myfont}
\newcommand{\logic}[1]{\text{\myfontcommand{#1}}}

\DeclareMathOperator{\tr}{tr}
\DeclareMathOperator{\Var}{Var}

\newcommand{\be}{\begin{equation}}
\newcommand{\ee}{\end{equation}}
\newcommand{\bea}{\begin{eqnarray}}
\newcommand{\eea}{\end{eqnarray}}
\newcommand{\bes}{\begin{equation*}}
\newcommand{\ees}{\end{equation*}}
\newcommand{\beas}{\begin{eqnarray*}}
\newcommand{\eeas}{\end{eqnarray*}}

\newcommand{\e}{\text{e}}
\newcommand{\ti}{\text{i}}

\begin{document}

\title{Quantum Optical Metrology of Correlated Phase and Loss}

\author{Patrick M. Birchall}
\affiliation{Quantum Engineering Technology Labs, H. H. Wills Physics Laboratory and Department of Electrical \& Electronic Engineering, University of Bristol, BS8 1FD, United Kingdom.}
\author{Euan J. Allen}
\email{euan.allen@bristol.ac.uk}
\affiliation{Quantum Engineering Technology Labs, H. H. Wills Physics Laboratory and Department of Electrical \& Electronic Engineering, University of Bristol, BS8 1FD, United Kingdom.}
\affiliation{Quantum Engineering Centre for Doctoral Training, H. H. Wills Physics Laboratory and Department of Electrical \& Electronic Engineering, University of Bristol, Tyndall Avenue, BS8 1FD, United Kingdom.}
\author{Thomas M. Stace}
\affiliation{ARC Centre of Excellence for Engineered Quantum Systems, School of Mathematics and Physics,
University of Queensland, Saint Lucia, Queensland 4072, Australia.}
\author{Jeremy L. O'Brien}
\affiliation{Quantum Engineering Technology Labs, H. H. Wills Physics Laboratory and Department of Electrical \& Electronic Engineering, University of Bristol, BS8 1FD, United Kingdom.}
\author{Jonathan C. F. Matthews}
\affiliation{Quantum Engineering Technology Labs, H. H. Wills Physics Laboratory and Department of Electrical \& Electronic Engineering, University of Bristol, BS8 1FD, United Kingdom.}
\author{Hugo Cable}
\affiliation{Quantum Engineering Technology Labs, H. H. Wills Physics Laboratory and Department of Electrical \& Electronic Engineering, University of Bristol, BS8 1FD, United Kingdom.}

\begin{abstract} 

Optical absorption measurements characterize a wide variety of systems from atomic gases to \emph{in-vivo} diagnostics of living organisms. Here we study the potential of non-classical techniques to reduce statistical noise below the shot-noise limit in absorption measurements with concomitant phase shifts imparted by a sample.  We consider both cases where there is a known relationship between absorption and a phase shift, and where this relationship is unknown.  For each case we  derive the fundamental limit and provide a practical strategy to reduce statistical noise.  Furthermore, we find an intuitive correspondence between measurements of absorption and of lossy phase shifts, which both show the same scope for precision enhancement. Our results demonstrate that non-classical techniques can aid real-world tasks with present-day laboratory techniques.

\end{abstract}
\date{\today}

\maketitle

The precision of optically measuring an object is limited by fundamental fluctuations in the optical field due to the statistical quantum nature of light~\cite{giovannetti2004}. When using laser light as an optical probe, the limit of this statistical noise is the shot-noise limit which can be reduced by increasing probe intensity or by enhancing interaction with the sample. However, some systems are incompatible with increased intensities, for example if light causes undesired technical effects~\cite{stace2010quantum,truong2015accurate} or the sample to deform~\cite{taylor2016,aasi2013}. If high-intensity light cannot be used then shot-noise will limit the achievable precision~\cite{stace2010quantum,truong2015accurate,kukura2010}. 

Whilst of a fundamental origin, shot-noise is not the ultimate quantum limit --- non-classical probes can be used to exceed the shot-noise limit~\cite{caves1981}. Many previous theoretical and experimental studies have investigated potential benefits of using non-classical states for phase estimation in the presence of loss~\cite{dorner2009,demkowicz2009,kacprowicz2010, thomas2011,cable2010,escher2011,demkowicz2013}, and for loss estimation~\cite{yurke1987,polzik1992,monras2007,adesso2009,brida2010,kacprowicz2010,aasi2013,sabines2016,moreau2016}. 
 In addition, a number of studies have investigated quantum bounds for multiparameter estimation including unitary (phase)~\cite{liu2017quantum,gessner2018sensitivity,proctor2018multiparameter} and non-unitary (phase, loss, de-phasing)~\cite{szczykulska2016multi,ragy2016compatibility,nichols2018multiparameter} channels.

At a fundamental level, changes in absorption over a narrow spectral range must be accompanied by changes in refractive index (and hence phase shifts), as governed by the Kramers-Kronig relations~\cite{libbrecht2006interferometric}. It is therefore important to consider how the estimation capabilities of any strategy are affected by correlation between these two variables. Here we address this and seek a unified understanding of quantum strategies for measuring absorption and phase of a single mode. Specifically, we consider estimating an unknown parameter $\chi$, which governs both phase $\theta(\chi)\!\in\![0,2\pi)$ and loss $1\!-\!\eta(\chi)\!\in\![0,1]$ imparted by a channel $\Lambda_\chi$ which we call correlated phase and loss estimation (CPLE). Formally, $\Lambda_{\chi}$ is defined by its action on a basis of coherent probe states $|\alpha\rangle\!\overset{\Lambda}{\mapsto}\!|\sqrt{\eta}\e^{\ti\theta}\alpha\rangle$. Lossy-phase estimation ($\partial_\chi\eta\!=\!0$ where $\partial_\bullet\!\equiv\!\frac{\partial}{\partial\bullet}$) and loss estimation ($\partial_{\chi}\theta\!=\!0$)~\cite{polzik1992,monras2007,adesso2009} are special cases of CPLE. 

We first find the fundamental upper bound on the precision achievable with CPLE, which is quantified using the quantum Fisher information (QFI) per input photon. We investigate the saturability of this bound using squeezed coherent states, which can readily be generated experimentally~\cite{andersen201630}. We also consider direct absorption estimation (DAE), where $\eta(\chi)$ is to be estimated but its relationship to $\theta(\chi)$ is not known and therefore the information contained in the phase cannot be accessed. We conclude by investigating multi-pass strategies for CPLE and DAE, and by investigating the advantage attainable in all cases by current experimental capabilities.

\textit{Fundamental limit for CPLE ---} We use the established Fisher information (FI) formalism to provide bounds on precision for estimating an unknown parameter $\chi$ encoded within a quantum state $\varrho_{\chi}$:
\begin{equation*}
\frac{1}{\Var(\chi)}\overset{1}{\leq}F^{\chi}_{\mathbf{M}}(\varrho_{\chi})\overset{2}{\leq}\mathcal{F}^{\chi}(\varrho_{\chi}).
\end{equation*}
Inequality 1 is the Cr\'{a}mer--Rao bound (CRB)~\cite{van2000} and relates the variance of unbiased estimates $\Var(\chi)$ to the FI $F^{\chi}_{\mathbf{M}}(\varrho_{\chi})\!=\!\sum_i p(i|\chi)\left[\partial_\chi\!\log{p}(i|\chi)\right]^2$. The FI is a function of the probabilities $p(i|\chi)\!=\!\tr(m_i\varrho_{\chi})$, given by the measurement of $\varrho_{\chi}$, with a positive-operator valued measure (POVM) $\mathbf{M} \!=\! \{m_i\}$ and $\sum_i m_i \!=\! \mathbb{1}$. Inequality 2 is the quantum CRB~\cite{helstrom1976} which relates  $F^{\chi}_{\mathbf{M}}(\varrho_{\chi})$ to its maximum value $\mathcal{F}^\chi$ (the QFI) which is found by optimizing over all POVMs~\cite{braunstein1994}. $\mathcal{F}$ serves as a measurement basis independent evaluation of the information that $\varrho_{\chi}$ contains on $\chi$. When $\chi$ is encoded onto a pure probe state by unitary $\mathcal{U}_{\chi}|\psi\rangle\!=\!|\psi^{\chi}\rangle$ the QFI becomes $4\left(\||\partial_\chi\psi^{\chi}\rangle\|^2\!-\!|\langle\psi^{\chi}|\partial_\chi\psi^{\chi}\rangle|^2\right)$ where $|\partial_\bullet\psi\rangle\!\equiv\!\partial_\bullet|\psi\rangle$ and $\|\bullet\|$ is the 2-norm.

Loss enacts a non-unitary evolution. Ref.~\cite{escher2011} showed that for such a non-unitary map $\Lambda_\chi$ acting on a pure state $|\psi\rangle$,
\begin{equation}\label{eq:minimisation}
\mathcal{F}[\Lambda_\chi(|\psi\rangle)]=\underset{\mathcal{U}_{\chi}}{\min}\,\big( \mathcal{F}\left[\mathcal{U}_{\chi}|\psi\rangle_S|0\rangle_E\right] \big),
\end{equation} 
where $\mathcal{U}_{\chi}$ is a unitary dilation of the channel, acting on a larger Hilbert space containing system mode $S$ and environment mode $E$, and satisfying $\Lambda_\chi(\bullet)=\tr_{E}\left[\mathcal{U}_{\chi}(\bullet_S\otimes|0\rangle\!\langle0|_E)\mathcal{U}_{\chi}^{\dagger}\right]$. For lossy-phase estimation $\mathcal{U}_{\chi}$ can be chosen such that $\mathcal{F}\left[ \mathcal{U}_{\chi}|\psi\rangle_S|0\rangle_E \right]$ provides informative bounds on the achievable precision dependent only on the mean number of probe photons $\langle\hat{n}\rangle_{\text{in}}\equiv\langle\psi|\hat{n}_S|\psi\rangle$~\cite{escher2011}.

Seeking an upper bound on the precision for CPLE we choose a unitary dilation of $S$, with a single free enviromental parameter $\varsigma$ which dictates the phase imparted onto $E$. This dilation takes the form $\mathcal{U}_{\chi,\varsigma}\!=\!U_2(\theta,\varsigma)U_1(\eta)$ where $U_1(\eta)$ and $U_2(\theta,\varsigma)$ enact system loss $(1\!-\!\eta)$ and phase $\theta$ of $\mathcal{U}$ respectively. These unitaries are given by $U_1\!=\!\exp[\ti \hat{H}_1\xi(\eta)]$, $\hat{H}_1\!=\!\frac{\ti}{2}(\hat{a}^{\dagger}_{S}\hat{a}_{E}\!-\!\hat{a}^{\dagger}_{E}\hat{a}_{S})$, $\xi(\eta)\!=\! \arccos(2\eta\!-\!1)$ and $U_2\!=\!\exp[\ti\hat{H}_2(\varsigma)\theta]$, $\hat{H}_2(\varsigma)\!=\!\hat{n}_S\!+\!\varsigma\hat{n}_E$. We verify that $\mathcal{U}_{\chi}$ is a dilation of $\Lambda_{\chi}$ in Supplementary Material A~\cite{supplmaterial}. In Supplementary Material B~\cite{supplmaterial} we show that for $|\Psi_{\chi,\varsigma}\rangle\equiv\mathcal{U}_{\chi,\varsigma}|\psi\rangle_S|0\rangle_E$:
\begin{equation} \label{qfisum}
\begin{split}
\mathcal{F}^{\chi}(|\Psi_{\chi,\varsigma}\rangle)&=
(\partial_{\chi} \theta)^24\left(\left\||\partial_\theta\Psi_{\chi,\varsigma}\rangle\right\|^2-|\langle\Psi_{\chi,\varsigma}|\partial_\theta\Psi_{\chi,\varsigma}\rangle|^2\right)\\
&+(\partial_\chi\eta)^24\left(\|\left|\partial_\eta\Psi_{\chi,\varsigma}\rangle \right\|^2-|\langle\Psi_{\chi,\varsigma}|\partial_\eta\Psi_{\chi,\varsigma}\rangle |^2\right).
\end{split}
\end{equation}
For any probe state, the second term in Eq.~(\ref{qfisum}) is given by:
\begin{equation}\nonumber
\begin{split}
(\partial_\chi\eta)^24\left(\left\||\partial_\eta\Psi_{\chi,\varsigma}\rangle\right\|^2-|\langle\Psi_{\chi,\varsigma}|\partial_\eta\Psi_{\chi,\varsigma}\rangle|^2\right)=\frac{(\partial_\chi\eta)^2\langle\hat{n}\rangle_{\text{in}}}{\eta(1-\eta)},
\end{split}
\end{equation}
which is independent of $\varsigma$~\cite{supplmaterial}. Therefore, the optimal $\varsigma$ is given by minimization of 
$\||\partial_\theta\Psi_{\chi,\varsigma}\rangle\|^2\!-\!|\langle\Psi_{\chi,\varsigma}|\partial_\theta\Psi_{\chi,\varsigma}\rangle |^2$, in accordance with Eq.~(\ref{eq:minimisation}).
This same expression was minimized in Ref.~\cite{escher2011} and hence has the same optimal value: 
\begin{equation}\label{eq:optimalvarsigma}
\varsigma_{\text{opt}}\!=\!1\!-\!\Var_{\text{in}}(\hat{n})\big/\left[(1\!-\!\eta)\Var_{\text{in}}(\hat n)\!+\!\eta\langle\hat n\rangle_{\text{in}}\right],
\end{equation}
with $\Var_{\text{in}}(\hat{n})\!=\!\langle\psi|{{\hat n}_S}^2|\psi\rangle\!-\!(\langle\hat{n}\rangle_{\text{in}})^2$. Therefore the limit we have found for CPLE is simply the sum of the limits on QFI for phase estimation (first term) and loss estimation (second term)~\cite{moreau2016}. Inserting $\varsigma_{\text{opt}}$ (Eq.~(\ref{eq:optimalvarsigma}))
into Eq.~(\ref{qfisum}) yields:
\begin{equation} \label{bound}
\begin{split}
\mathcal{F}^{\chi}(\varrho_{\chi})&\leq(\partial_\chi\theta)^2 \bigg[\frac{4\eta \langle \hat n\rangle_{\text{in}} \Var_{\text{in}}(\hat n)}{(1-\eta)\Var_{\text{in}}(\hat{n})+\eta\langle\hat{n}\rangle_{\text{in}}}\bigg] \\
&+(\partial_\chi\eta)^2\frac{\langle\hat{n}\rangle_{\text{in}}}{\eta(1-\eta)} \\
&\leq\langle\hat{n}\rangle_{\text{in}}\frac{4\eta^2(\partial_\chi\theta)^2+(\partial_\chi\eta)^2}{\eta(1-\eta)} \eqqcolon \mathcal{Q}_{\chi}.
\end{split}
\end{equation}
where the last expression depends only on $\langle\hat{n}\rangle_{\text{in}}$. $\mathcal{Q}_{\chi}$ denotes the maximum information available on $\chi$ for any quantum probe and measurement, and therefore the bound we aim to saturate. We note that phase estimation benefits from super-Poissionian statistics in pure states, $\Var_{\text{in}}(\hat{n}) \geq \langle\hat{n}\rangle_{\text{in}}$ \cite{braunstein1994}, while loss estimation benefits from sub-Poissionian statistics, $\Var_{\text{in}}(\hat{n})\leq\langle\hat{n}\rangle_{\text{in}}$~\cite{adesso2009} --- however, a probe state cannot have both properties. This suggests the inequality in Eq.~(\ref{bound}) may not be saturable. However, we show that this bound can be saturated.

\textit{Probe states for CPLE ---} Having found the fundamental limit for CPLE, we next seek an effective strategy for experimentally achieving this bound using single-mode Gaussian states and homodyne measurements. These were recently shown to be optimal for lossy-phase estimation in the large photon number limit~\cite{birchall2017quantum}. 

Single-mode Gaussian states are specified by a displacement vector $\bm{d}$ comprised of means, $d_i\!=\!\langle\hat{x}_i\rangle$, and a matrix $\mathbf{\Gamma}$ comprised of covariances, $\Gamma_{ij}\!=\!\frac{1}{2}\langle\hat{x}_i\hat{x}_j\!+\!\hat{x}_j\hat{x}_i\rangle\!-\!\langle\hat{x}_i\rangle\langle\hat{x}_j\rangle$, of the quadrature operators $\hat{x}_1\!=\!\frac{1}{2}(\hat{a}^{\dagger}\!+\!\hat{a})$ and $\hat{x}_2\!=\!\frac{1}{2}\ti(\hat{a}^{\dagger}\!-\!\hat{a})$ \cite{simon1987,simon1994}. Homodyne measurement of a single-mode state provides a measurement of the $\hat{x}_1$ quadrature \cite{loudon1987}. An arbitrary single-mode pure Gaussian state can be defined by the squeezing $\hat{S}(r,\phi)\!=\!\exp[\frac{1}{2}r({\e}^{-\ti\phi}\hat{a}^2\!-\!{\e}^{\ti\phi}{\hat{a}^{\dagger2}})]$, displacement $\hat{D}(\alpha)\!=\!\exp[\alpha(\hat{a}^{\dagger}\!-\!\hat{a})]$, and rotation $\hat{R}(\varphi)\!=\!\exp(\ti\hat{a}^{\dagger}\hat{a}\varphi)$ operators acting on vacuum:
$|\psi^{\text{G}}\rangle=\hat{R}(\varphi)\hat{D}(\alpha)\hat S(r,\phi)|0\rangle$
where all arguments are real and the mean number of photons within the state is: $\langle\hat{n}\rangle\!=\!\alpha^2+\sinh^2(r)$. The actions of squeezing, displacement, rotation (phase shift) and loss modify $\bm{d}$ and $\mathbf{\Gamma}$~\cite{weedbrook2012}. 
$|\psi^{\text{G}}\rangle$ will be transformed by $\Lambda_\chi$ to $\tilde\varrho\!=\!\Lambda_\chi(|\psi^{\text{G}}\rangle)$ with $\tilde{\bm{d}}\!=\!\mathcal{R}(\varphi\!+\!\theta)\!\begin{pmatrix}\alpha\sqrt{\eta}\\ 0\end{pmatrix}$ and  $\tilde{\mathbf{\Gamma}} =\! \mathcal{R}\!(\varphi \!+\! \phi/2 \!+\! \theta)\displaystyle{\frac{1}{4}}\!\begin{pmatrix}\eta\e^{-2r}\!+\!1\!-\!\eta&\hspace{-15pt}0\hspace{-15pt}\\\hspace{-15pt}0&\hspace{-15pt}\eta\e^{2r}\!+\!1\!-\!\eta\end{pmatrix}\!\mathcal{R}^{\top}\!(\varphi\!+\!\phi/2\!+\!\theta)$, where $\mathcal{R}(\bullet)\!=\!\begin{pmatrix}\cos\bullet&-\sin\bullet\\\sin\bullet&\cos\bullet\end{pmatrix}$ is the rotation matrix~\cite{loudon1987}. Throughout the following, tildes over variables refer to properties of the state after $\Lambda_\chi$ has been applied. $\bm{d}$ and $\mathbf{\Gamma}$ of $|\psi^{\text{G}}\rangle$ can be observed by setting $\eta\!=\!1$ and $\theta\!=\!0$ in $\tilde{\bm{d}}$ and $\tilde{\mathbf{\Gamma}}$.

The QFI of a single-mode Gaussian state $\tilde\varrho$ is~\cite{pinel2013}:
\begin{equation} \label{qfism}
\mathcal{F}^{\chi}(\tilde\varrho)=\frac{\tr[({\tilde{\mathbf{\Gamma}}}^{-1}\partial_{\chi}\tilde{{\mathbf{\Gamma}}})^2]}{2(1+{\tilde{P}}^2)}+\frac{2(\partial_{\chi}\tilde{P})^2}{1-\tilde{P}^4}+(\partial_\chi\tilde{\bm{d}})^\top\tilde{\mathbf{\Gamma}}^{-1}(\partial_\chi\tilde{{\bm{d}}}),
\end{equation}
where $\tilde{P}\!=\!\tr(\tilde{\varrho}^2)$ is the purity. Directly optimising the QFI of a Gaussian state for lossy-phase estimation provides sub-optimal use with homodyne measurement~\cite{aspachs2009}. Because of this, we optimize information related to the parameter dependence on displacement vector $\tilde{\bm{d}}$, in the third term of Eq.~(\ref{qfism}). For lossy-phase estimation it was shown that this information is accessible through homodyne detection and thus we seek to maximise this term by varying the probe $|\psi^{\text{G}}\rangle$. 

To do this, the squeezing angle $\phi$ should be set such that $\partial_\chi\tilde{{\bm{d}}}$ is parallel to the direction of minimum uncertainty in the output state i.e. aligned with the eigenvector of $ \tilde{\mathbf{\Gamma}}$ with smallest eigenvalue $\tilde{V}_{\text{min}}\!=\![\e^{-2r}\eta\!+\!(1\!-\!\eta)]/4$. A state satisfying this condition is plotted in Fig.\ref{phase}. In this case, the information contained in displacement vector $\mathcal{D}$ is given by 
\begin{equation} 
\label{Ddefinition}
\mathcal{D}\!\coloneqq\!(\partial_{\chi}\tilde{\bm{d}})^\top\tilde{\mathbf{\Gamma}}^{-1}(\partial_\chi\tilde{{\bm{d}}})\!=\!\|\partial_\chi\tilde{\bm{d}}\|^2\big/\tilde{V}_{\text{min}}.
\end{equation}
The output can be measured using homodyne detection to produce a signal which has a FI of $\mathcal{D}\!+\!(\partial_\chi\tilde{V}_{\text{min}})^2/(2{\tilde{V}_{\text{min}}}^2)$~\cite{van2000}, which shows that $\mathcal{D}$ is a quantity which can be accessed with a practical measurement. Using an adaptive feedback strategy (e.g.~\cite{berni2015}), the squeezing and homodyne angles can be set arbitrarily close to their optimal values. $\partial_\chi{\tilde{\bm{d}}} = (\partial_\chi\theta)\partial_\theta\tilde{\bm{d}}+(\partial_\chi\eta)\partial_\eta \tilde{\bm{d}}$ where the two terms are always orthogonal, therefore:
\begin{figure}
\includegraphics[width=0.59\columnwidth]{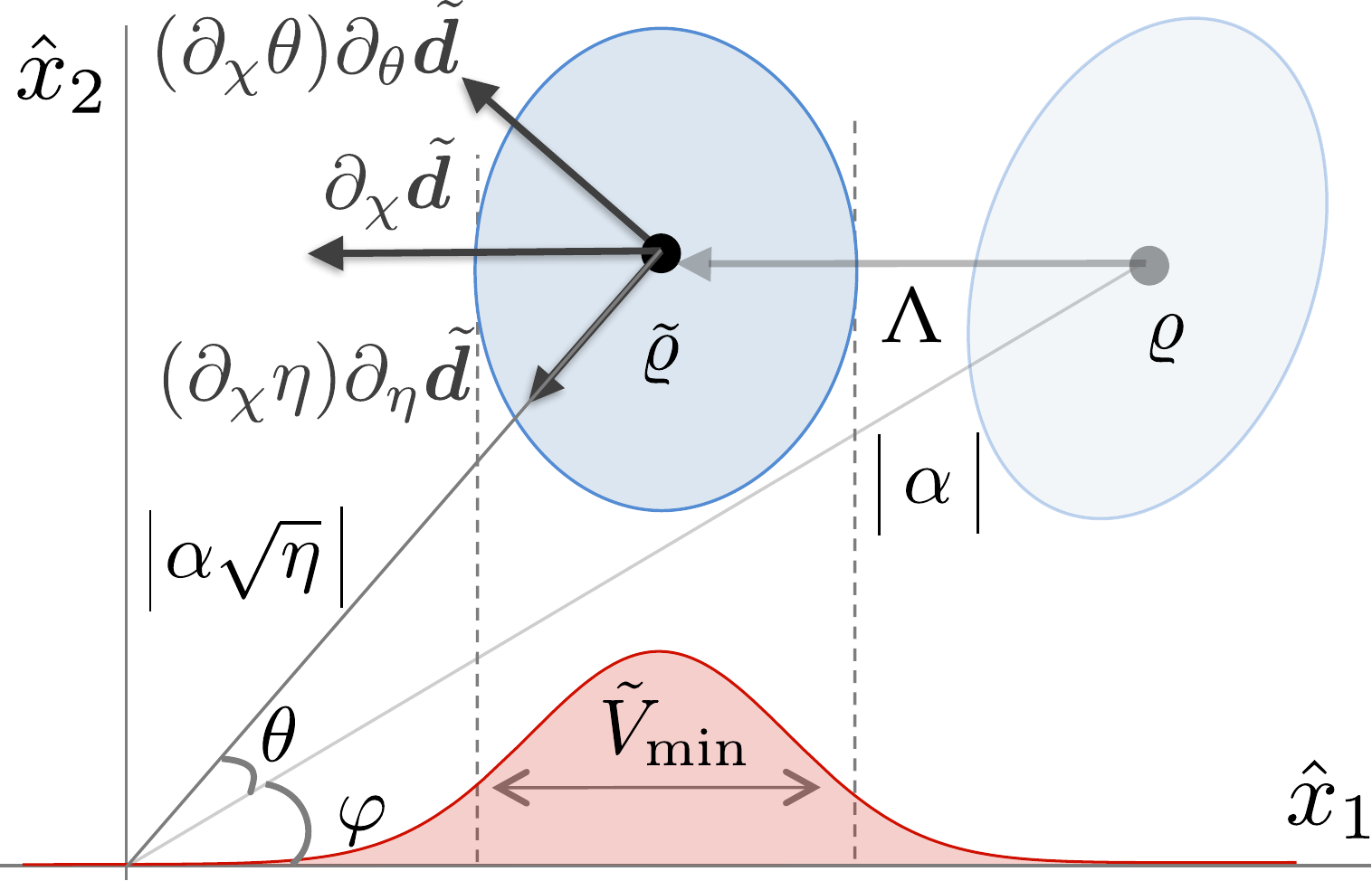}
\caption{\textbf{Phase-space representation of the transformation of initial state $\varrho$ to $\tilde\varrho$}: after passing through the channel $\Lambda$ with transmission $\eta$ and a phase shift of $\theta$. $\varrho$ is squeezed in the optimal direction aligned with $\partial_{\chi}\tilde{\bm{d}}$. The red curve is the homodyne signal when the phase of the local-oscillator is optimized for the measurement.}
\label{phase}
\end{figure}
\begin{figure*}
\includegraphics[width=\textwidth]{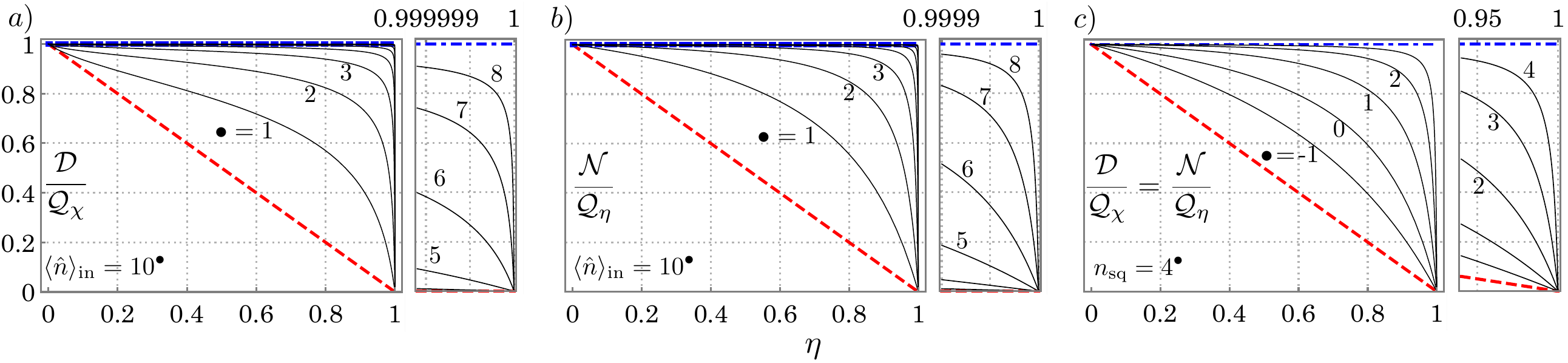}
\caption{\textbf{Comparing strategies for CPLE and DAE with their respective quantum limits:} Within each plot the red dashed line shows the SQL. a) Amount of statistical information $\mathcal{D}$ encoded onto the displacement vector (mean number of photons) of a squeezed coherent state for CPLE , normalised to the quantum limit $\mathcal{Q}_{\chi}$. The inset shows that even for very low absorption these states approach the quantum limit for modest energies. Statistical information plotted for varying input mean photon number operating with the optimal squeezing value presented in Eq.~(\ref{optimalSqueezing}). b) Amount of statistical information $\mathcal{N}$ encoded onto the displacement vector (mean number of photons) of a squeezed coherent state for DAE, normalised to the quantum limit $\mathcal{Q}_{\eta}$. Statistical information plotted for varying input mean photon number operating with the optimal squeezing value presented in Eq.~(\ref{optimalSqueezing}). c) Amount of statistical information $\mathcal{D}$ ($\mathcal{N}$) encoded onto the displacement vector (mean number of photons) of a squeezed coherent state for CPLE (DAE), normalised to the quantum limit $\mathcal{Q}_{\chi}$ ($\mathcal{Q}_{\eta}$) when $\alpha$ is large.}\label{threeplots}
\end{figure*}
\begin{equation*}
\begin{split}
\|\partial_\chi\tilde{\bm{d}}\|^2&=\|(\partial_{\chi}\theta)\partial_{\theta}\tilde{\bm{d}}\|^2+\|(\partial_\chi\eta)\partial_\eta\tilde{\bm{d}}\|^2\\
&={\alpha^2[4\eta^2(\partial_\chi\theta)^2+(\partial_\chi\eta)^2]}/{4\eta},
\end{split}
\end{equation*}
where $\alpha$ is the coherent amplitude of the input state (Fig.~\ref{phase}) and $(\partial_\chi\theta)^2$ and $(\partial_\chi\eta)^2$ appear in the same proportions as in $\mathcal{Q}_\chi$ (Eq.~(\ref{bound})). It can be observed from Eq.~(\ref{qfism}) that the QFI achieved with an unsqueezed coherent state as probe is
\begin{equation} \label{classical_cple}
\mathcal{D}\big|_{r=0}={\langle\hat{n}\rangle_{\text{in}}[4\eta^2(\partial_\chi\theta)^2+(\partial_\chi\eta)^2]}/{\eta}\coloneqq\mathcal{S}_{\chi},
\end{equation}
which limits the best precision achievable using classical probes with a single pass through $\Lambda_\chi$ --- the standard quantum limit (SQL). 

Combining $\tilde{V}_{\text{min}}$ and Eq.~(\ref{Ddefinition}) we find
\begin{equation}\label{dprime}
\mathcal{D}=(\langle\hat{n}\rangle_\text{in}-n_{\text{sq}})\frac{4\eta^2 (\partial_\chi\theta)^2+(\partial_\chi\eta)^2}{\eta\left[\e^{-2r}\eta+(1-\eta)\right]}.
\end{equation}
where $\alpha^2\!=\!\langle\hat{n}\rangle_{\text{in}}-n_{\text{sq}}$ has been used and $n_{\text{sq}}\!=\!\sinh^2(r)$ is the number of photons contributing to the squeezing of the input state. As $\langle\hat{n}\rangle_{\text{in}}$ grows, $\mathcal{D}/\langle\hat{n}\rangle_{\text{in}}$ will converge to the quantum limit we have found in Eq.~(\ref{bound}) i.e. $\lim_{\langle\hat{n}\rangle_\text{in}\rightarrow\infty}\mathcal{D}/ \langle\hat{n}\rangle_{\text{in}}\!=\!\mathcal{Q}_\chi/ \langle\hat{n}\rangle_{\text{in}}$ if two conditions are satisfied: First, $n_{\text{sq}}$ needs to be a vanishing proportion of the total number of probe photons $\lim_{\langle\hat{n}\rangle_{\text{in}}\rightarrow\infty}n_{\text{sq}}/ \langle\hat{n}\rangle_{\text{in}}\!=\!0$. Second, $n_{\text{sq}}$ needs to be unbounded with increasing $\langle\hat{n}\rangle_{\text{in}}$, which will ensure $\e^{-2r}$ vanishes. In Supplementary Material C~\cite{supplmaterial} we describe a state with finite, and arbitrary, $\langle\hat{n}\rangle_{\text{in}}$ for which $\mathcal{F}^\chi(\varrho)\!=\!\mathcal{Q}_{\chi}$, demonstrating $\mathcal{Q}_{\chi}$ is a saturable upper bound (though not of genuine practical utility).

Therefore, we have found that there is no trade-off in the information encoded on a state by the phase and loss of a channel. This is in contrast to the task of estimating phase and loss when there is no correlation~\cite{crowley2014} which displays a necessary trade-off in the precision to which each parameter could be estimated. Our results also contrast with those reported in Ref.~\cite{dinani2016}, which assume total energy of a probe state including any reference or ancilla (which does not expose the sample) as the resource.  With this assumption it was found for the low photon-number regime that there is a trade-off in the sensitivity of the probe state to either  loss or phase. Our choice of resource (the total optical power incident on the sample) is relevant when the sample is delicate. The total optical power in a probe often constitutes a small fraction of the total energy needed for example to generate the quantum probe.  
 
For finite $\langle\hat{n}\rangle_{\text{in}}$, $\mathcal{D}$ can be optimised by choosing the best value of $n_{\text{sq}}$. The optimal amount of squeezing is derived in the Supplementary Material D~\cite{supplmaterial} to be
\begin{equation}\label{optimalSqueezing}
n_{\text{sq}}= \frac{\left(\sqrt{1-4 (\eta -1) \eta  \langle\hat{n}\rangle_{\text{in}}}-1\right)^2}{4 (1-\eta) \left(\sqrt{1-4 (\eta -1) \eta  \langle\hat{n}\rangle_{\text{in}}}-\eta \right)}, 
\end{equation} 
which results in
\begin{equation*}
\mathcal{D} =\mathcal{Q}_\chi\,\frac{2 (\eta -1) \langle\hat{n}\rangle_{\text{in}}+\sqrt{1-4 (\eta -1) \eta  \langle\hat{n}\rangle_{\text{in}}}-1}{2 (\eta -1) \langle\hat{n}\rangle_{\text{in}}}.
\end{equation*} 
In Fig.~\ref{threeplots}a the optimal $\mathcal{D}$ for a selection of different values of $\langle\hat{n}\rangle_{\text{in}}$ is plotted over $\eta\!\in\!(0,1)$. The range of $\langle\hat{n}\rangle_{\text{in}}\!=\!10^i,i\!\in\!\{0,1,...,8\}$ scale to large numbers but corresponds to low energy e.g.~$10^8$ photons at $\lambda\! =\!500$~nm equates to $4\times\!10^{-11}$ J. The plot shows that Gaussian states with modest energies can provide large precision gains for CPLE.

\textit{Probe states for DAE ---} 
We now turn to DAEs, which refer to measurements of absorption which do not exploit information about any phase imparted by a sample. Previously, a limit on QFI was found for transmission estimation where no phase is imparted by the sample i.e. $\theta = 0$ \cite{monras2007}, and Fock states were identified as optimal for this \cite{adesso2009}. This bound applies equally for DAE since Fock states are invariant under phase shifts. Since $\theta$ is uncorrelated with $\eta$ and unknown, it cannot increase the QFI associated with $\eta$ \cite{helstrom1976}, and therefore the limit on QFI for DAE is $\mathcal{Q}\displaystyle{\big|_{\partial_\chi\theta=0,\,\partial_\chi\eta=1}}\coloneqq\mathcal{Q}_{\eta}$. Similarly $\mathcal{S}_{\chi}\displaystyle{\big|_{\partial_\chi\theta = 0,\,\partial_\chi\eta=1}}\coloneqq\mathcal{S}_{\eta}$, is the SQL for DAE \cite{adesso2009}. However when a Gaussian probe is used, DAE is inequivalent to CPLE with $\partial_\chi\theta=0$ since the probe state will be transformed by any phase shift present. For instance, the strategy for CPLE described above using Gaussian states does not work for DAE as the correct homodyne measurement setting depends on the phase imparted by the sample --- we therefore seek an alternative strategy.

Intensity measurements are unaffected by the phase of the detected light, and therefore provide a way to decouple the effects of sample absorption and any phase shift. To find useful strategies for DAE, we consider the statistical information~$\mathcal{N}$ contained measurement of the mean intensity which will be detected $\langle\hat{n}\rangle_{\text{out}}\!=\!\eta\langle\hat{n}\rangle_{\text{in}}$, which can be found most simply using standard error propagation:
\begin{equation}\label{bign}
\begin{split}
\mathcal{N} \coloneqq 1/\Var(\eta)&=\left(\partial_\eta\langle\hat{n}\rangle_{\text{out}}\right)^2\big/\Var_{\text{out}}({\hat{n}})  \\
&=(\langle\hat{n}\rangle_{\text{in}})^2\big/ \left[\eta^2\Var_{\text{in}}({\hat{n}})+\eta(1-\eta)\langle\hat{n}\rangle_{\text{in}}\right],
\end{split}
\end{equation}
which applies for arbitrary states. Considering only the mean intensity ensures complex measurement and estimation procedures are not needed and $\mathcal{N}$ plays a role analogous to FI.  

Loss reduces the amplitude of a Gaussian state, and so a natural probe state to consider for DAE is an amplitude-squeezed Gaussian state, $|\psi^\text{G}\rangle\big|_{\phi=0}$. Noting that $\Var_{\text{in}}({\hat n})\!=\!2\tr^2\mathbf{\Gamma}\!-\frac{3}{4}+(\langle\hat{n}\rangle_{\text{in}}-n_\text{sq})\e^{-2r}$ for an amplitude squeezed Gaussian state \cite{dodonov1994}, and $\tr^2 \mathbf{\Gamma}\!=\!\mathcal{O}(n_{\text{sq}}^2)$. Asymptotic optimality  $\lim_{\langle\hat{n}\rangle_\text{in}\rightarrow\infty}\mathcal{N}/ \langle\hat{n}\rangle_{\text{in}}\!=\!\mathcal{Q}_\eta/ \langle\hat{n}\rangle_{\text{in}}$ can be achieved if $n_{\text{sq}}$ is unbounded (to ensure $\e^{-2r}$ vanishes) and also a vanishing proportion of $\sqrt{\langle\hat{n}\rangle_{\text{in}}}$. This ensures that the photon number variance of the input state contributes negligibly to the denominator of expression on the second line of Eq.~(\ref{bign}). Also shown in Eq.~(\ref{bign}) is that in order to maximize $\mathcal{N}$, the photon number variance of the input state should be minimised for a given $\langle\hat{n}\rangle_{\text{in}}$ independently of $\eta$. In Fig.~\ref{threeplots}b the optimal $\mathcal{N}$ for a selection of different values of $\langle\hat n\rangle_{\text{in}}$ is plotted over $\eta\!\in\!(0,1)$. (see Supplementary Material E~\cite{supplmaterial} for the optimization). This plot shows that Gaussian states with modest energies can provide large precision gains for DAE.

\textit{Multi-pass strategies ---} Rather than using non-classical states, it is sometimes possible to increase precision beyond the SQL by sending a classical (coherent state) optical probe through the sample multiple times~\cite{birchall2017quantum}. Recently it was shown that, for lossy-phase estimation, multi-pass strategies could obtain 60$\%$ of the quantum limit on FI for a given number of photons incident upon the sample over all passes and for any values of the phase shift and loss. In Supplementary Material F~\cite{supplmaterial} we extend this result and show that multi-pass strategies provide exactly the same benefits for CPLE and DAE as they do for lossy-phase estimation. This exact correspondence holds even when lossy components are used to perform the multi-pass strategy. 

\textit{Practical application ---} At present the highest amount of optical squeezing demonstrated is 15 dB~\cite{vahlbruch2016} ($n_{\text{sq}}\!=\!7.4$). By explicitly considering large $\alpha$ we can quantify the quantum advantage, $\Delta$, squeezing brings to both CPLE and DAE:
\begin{equation}\label{largealpha}
\Delta = \lim_{\alpha\rightarrow\infty}\mathcal{N}\big/\mathcal{S}_{\eta}=\lim_{\alpha\rightarrow\infty}\mathcal{D}\big/\mathcal{S}_\chi=\frac{1}{\e^{-2r}\eta+(1-\eta)},
\end{equation}
observing that the enhancement provided for both DAE and CPLE is the same. The precision gains which squeezing brings to probe states with large $\alpha$ is plotted in Fig.~\ref{threeplots}.c.

For CPLE, Eq.~(\ref{largealpha}) encouragingly indicates that a small amount of squeezing can substantially increase the precision of a measurement. Generating and detecting Fock states is a non-trivial task and as such only low photon number Fock states have been generated~\cite{cooper2013,motes2016}; these states may prove useful for the measurement of samples which are damaged by very few photons. The Gaussian probe state we have studied can be created by the displacement of a squeezed vacuum state to contain much larger amounts of power~\cite{schneider1996}, benefiting absorption measurements far beyond the few photon regime.

We highlight Ref.~\cite{kukura2010} which reported absorption measurements with 10~$\mu$W of incident laser light ($10^{13}$ photons per second) at 633 nm to detect the presence of single molecules. Using a balanced photodetector the effective intensity fluctuations in the laser light were reduced to the shot-noise limit. Using Eq.~(\ref{largealpha}) and taking $\eta$ to be $0.95$, 15 dB of squeezing in this experiment would reduce the contribution to the mean-squared error (MSE) from fundamental fluctuations by a factor of 12.5. This is 79\% of the advantage provided by using $10^{13}$ photons per second in Fock states. Alternatively, the same precision could be achieved with a factor of 12.5 reduction in input intensity.

\textit{Conclusion ---} Our results further indicate that for estimating parameters of linear optical transformations with non-unit transmissivity, the information which can be encoded in the coarse-grained properties of a state, such as the mean intensity or mean quadrature value, is very close to the fundamental limit on the information which can be encoded on an entire state~\cite{demkowicz2013,birchall2017quantum}. We anticipate the quantum limit on CPLE and our Gaussian state strategy can be generalized to multiparameter estimation problems~\cite{albarelli2019evaluating} and perhaps even to precision estimation of general-linear mode transformations.

\begin{acknowledgments}
\textit{Acknowledgments ---} We thank J.P. Dowling and D.H. Mahler for helpful discussions. This work was supported by EPSRC, ERC, PICQUE, BBOI, US Army Research Office (ARO) Grant No. W911NF-14-1-0133, U.S. Air Force Office of Scientific Research (AFOSR) and the Centre for Nanoscience and Quantum Information (NSQI). E.J.A. was supported by the Quantum Engineering Centre for Doctoral Training, EPSRC grant EP/L015730/1. J.L.O’B. acknowledges a Royal Society Wolfson Merit Award and a Royal Academy of Engineering Chair in Emerging Technologies. J.C.F.M. and J.L.O'B acknowledge fellowship support from EPSRC. J.C.F.M. acknowledges support from ERC starting grant ERC-2018-STG803665. T.M.S. was supported by the Benjamin-Meaker visiting fellowship and the ARC Centre of Excellence in Engineered Quantum Systems CE17.
\end{acknowledgments}


\bibliographystyle{ieeetr}

\begin{thebibliography}{10}
\bibitem{footnote1}
This is in accordance with the operator-norm based methods for finding upper
bounds in metrology of qubit systems, since linear optical transformations are
  separable in single particle representation wherein photons in a two-mode
  interferometer become qubits and separable dilations provide the tightest
  bounds when the number of qubits are large. However, as states with a fixed
  mean number of photons do not form a subspace, operator norm based methods
  cannot be directly applied.
\end{thebibliography}

\onecolumngrid
\newpage

\section{Supplementary Material A: Channel Dilation} \label{sec:channel_dilation}
Letting the system mode be mode one, $\hat{a}^{\dagger}_{S}= \hat{a}^{\dagger}_{1}$ and the environment mode be mode two, $\hat{a}^{\dagger}_{E}= \hat{a}^{\dagger}_{2}$ the unitary dilation stated in the main text is
\begin{equation}
\mathcal{U}_{\chi,\varsigma}=U_2(\theta,\varsigma)U_1(\eta)
\end{equation}
where $U_1=\exp[\ti \hat{H}_1\xi(\eta)]$, $\hat{H}_1=\frac{\ti}{2}(\hat{a}^{\dagger}_{1}\hat{a}_{2}-\hat{a}^{\dagger}_{2}\hat{a}_{1})$, $\xi(\eta)=2\arccos(\sqrt{\eta})$ and $U_2=\exp[\ti\hat{H}_2(\varsigma)\theta]$, $\hat{H}_2(\varsigma)=\hat{n}_1+\varsigma\hat{n}_2$. The transfer matrix associated with $U_1$ is:
\begin{equation}
\begin{split}
\e^{-\ti \xi(\eta) Y/2} &=
\begin{pmatrix}
\cos\left[\xi(\eta)/2\right] & -\sin\left[\xi(\eta)/2\right] \\
\sin\left[\xi(\eta)/2\right] & \cos\left[\xi(\eta)/2\right] \\
\end{pmatrix} \\
& =
\begin{pmatrix}
\sqrt{\eta} & -\sqrt{1-\eta} \\
\sqrt{1-\eta} & \sqrt{\eta} \\
\end{pmatrix}.
\end{split}
\end{equation}
where $Y = \begin{pmatrix}
0 & -\ti \\
\ti & 0 \\
\end{pmatrix} $ is the Pauli Y matrix. Similarly, the transfer matrix associated with $U_2$ is:
\begin{equation}
\begin{pmatrix}
\e^{\ti \theta} & 0 \\
0 & \e^{\ti \varsigma \theta} \\
\end{pmatrix}
\end{equation}
such that the transfer matrix of $\mathcal{U_{\chi,\varsigma}}$ is
\begin{equation}
\begin{pmatrix}
\e^{\ti \theta}\sqrt{\eta} & -\e^{\ti \theta}\sqrt{1-\eta} \\
\e^{\ti \varsigma \theta}\sqrt{1-\eta} & \e^{\ti \varsigma \theta}\sqrt{\eta} \\
\end{pmatrix}.
\end{equation}
The element $T_{11}$ is an attenuation by $\eta$ and a phase shift $\theta$ as desired.

\section{Supplementary Material B: Bounding the Quantum Fisher Information for Correlated Phase and Loss Estimation} \label{sec:separating_qfi}
In this section we continue to use the notation $\hat{a}^{\dagger}_{S}= \hat{a}^{\dagger}_{1}$ and  $\hat{a}^{\dagger}_{E}= \hat{a}^{\dagger}_{2}$ introduced in Section~A
. We start with an expression of the QFI for pure states:
\begin{equation} \label{pureqfi}
\mathcal{F}\left(|\Psi_{\chi,\varsigma}\rangle\right) = 4\left(\langle \partial_\chi{\Psi}_{\chi,\varsigma}|\partial_{\chi}{\Psi}_{\chi,\varsigma}\rangle- |\langle \Psi_{\chi,\varsigma} |\partial_\chi{\Psi}_{\chi,\varsigma}\rangle|^2 \right).
\end{equation}
Focusing on the second term in Eq.\,(\ref{pureqfi}):
\begin{equation}
\begin{split} \label{qfitermtwo}
\langle \Psi_{\chi,\varsigma} |\partial_\chi{\Psi}_{\chi,\varsigma}\rangle & = \langle \psi,0|\mathcal{U}^{-1}\left(\partial_{\chi}{\mathcal{U}}\right)|\psi,0\rangle \\
& = \langle \psi,0|\mathcal{U}^{-1}\left\{\ti\left[\left(\partial_\chi\theta\right) \hat H_2 + (\partial_{\chi}{\eta}) (\partial_{\eta}\xi)U_2\hat H_1U_2^{-1}\right]\mathcal{U}\right\}|\psi,0\rangle \\
& =\ti (\partial_\chi \theta)  \langle \psi,0|U_1^{-1}\hat H_2 U_1|\psi,0\rangle \\
& \hspace{10pt} +  \ti (\partial_\chi{\eta})(\partial_\eta \xi) \langle \psi,0|\hat H_1|\psi,0\rangle \\
& = \ti (\partial_\chi\theta) \langle \psi,0|U_1^{-1}\hat H_2 U_1|\psi,0\rangle
\end{split}
\end{equation}
where in the first step we used
\begin{equation}
\begin{split}
\partial_{\chi}{\mathcal{U}} & = (\partial_\chi U_2)U_1 +  U_2(\partial_\chi U_1) \\
& =  (\partial_\chi \theta)\,  \ti\, \hat H_2\, \e^{\ti \hat H_2 \theta} U_1 +  U_2(\partial_\chi \eta)(\partial_\eta \xi)\, \ti \, \hat H_1 \, \e^{\ti \hat H_1 \xi} \\
& = \ti\left[(\partial_\chi\theta) \hat H_2 + (\partial_\chi{\eta}) (\partial_\eta\xi) U_2\hat H_1U_2^{-1}\right]\mathcal{U} \\
\end{split}
\end{equation}
and in the second step we used $\langle \psi,0|\hat H_1|\psi,0\rangle  = 0$ which is due to the form of $\hat{H}_1=\frac{\ti}{2}(\hat{a}^{\dagger}_{1}\hat{a}_{2}-\hat{a}^{\dagger}_{2}\hat{a}_{1})$. Expanding the first term in Eq.\,(\ref{pureqfi}) gives:
\begin{equation} \label{qfitermone}
\begin{split}
\langle \partial_\chi{\Psi}_{\chi,\varsigma}|\partial_{\chi}{\Psi}_{\chi,\varsigma}\rangle & = \langle  \psi,0|\left(\partial_{\chi}\mathcal{U}^{-1}\right) \left(\partial_{\chi}{\mathcal{U}}\right)|\psi,0\rangle \\
& =(\partial_\chi \theta)^2 \langle \psi,0|U_1^{-1}\hat {H}_2^2 U_1|\psi,0\rangle  \\
& \hspace{10pt}+ \left[ (\partial_\chi\eta) (\partial_\eta \xi) \right]^2 \langle \psi,0|\hat H^2_1|\psi,0\rangle \\
& \hspace{10pt}+ (\partial_\chi\theta) (\partial_\chi\eta) (\partial_\eta \xi)  \langle \psi,0|\{\hat H_1,U_1^{-1} \hat H_2U_1 \} |\psi,0\rangle
\end{split}
\end{equation}
where $\{A,B\}=AB+BA$ is the commutator. By using
\begin{equation} \label{phase_loss_similarity}
\begin{split}
U_{1}^{-1}  \hat{a}^{\dagger}_{1} U_{1}& = \sqrt{\eta}  \hat{a}^{\dagger}_{1} - \sqrt{1-\eta}  \hat{a}^{\dagger}_{2} \\
U_{1}^{-1}  \hat{a}^{\dagger}_{2} U_{1}& = \sqrt{1- \eta}  \hat{a}^{\dagger}_{1} + \sqrt{\eta}  \hat{a}^{\dagger}_{2} \\
U_{1}^{-1}  \hat{a}_{1} U_{1}& = \sqrt{\eta}  \hat{a}_{1} - \sqrt{1-\eta}  \hat{a}_{2} \\
U_{1}^{-1}  \hat{a}_{2} U_{1}& = \sqrt{1- \eta}  \hat{a}_{1} + \sqrt{\eta}  \hat{a}_{2} \\
\end{split}
\end{equation}
we can calculate:
\begin{equation}
U_1^{-1}\hat H_2 U_1 = (\eta +\varsigma - \eta \varsigma )\, \hat n_1 + \eta(\varsigma-1)\sqrt{1-\eta}\,\hat n_2 + (\varsigma -1)\sqrt{\eta(1-\eta)} \left(\hat{a}^{\dagger}_{1}\hat{a}_{2} + \hat{a}^{\dagger}_{2}\hat{a}_{1} \right).
\end{equation}
As neither $\hat n_1$ nor the $\hat n_2$ operator will populate mode two with photons we can see that
\begin{equation}
\langle \psi, 0| \hat n_1 \hat H_1 |\psi, 0 \rangle = \langle \psi, 0| \hat n_2 \hat H_1 |\psi, 0 \rangle = \langle \psi, 0|  \hat H_1\hat n_1 |\psi, 0 \rangle = \langle \psi, 0|  \hat H_1\hat n_2 |\psi, 0 \rangle = 0
\end{equation}
therefore, letting $\gamma = (\varsigma -1)\sqrt{\eta(1-\eta)}$, we can evaluate the last term in Eq.~(\ref{qfitermone}):
\begin{equation} \label{crossterm}
\begin{split}
\langle \psi,0|\{\hat H_1,U_1^{-1} \hat H_2U_1 \} |\psi,0\rangle& = \gamma \langle \psi,0|\{\hat H_1,(\hat{a}^{\dagger}_{1}\hat{a}_{2} + \hat{a}^{\dagger}_{2}\hat{a}_{1}) \} |\psi,0\rangle \\
& = \gamma \ti  \langle \psi,0|\left[\left(\hat{a}^{\dagger}_{1}\hat{a}_{2} \right)^2 - \left(\hat{a}^{\dagger}_{2}\hat{a}_{1} \right)^2\right] |\psi,0\rangle \\
& = 0.
\end{split}
\end{equation}
Putting Eq.~(\ref{qfitermone}), Eq.~(\ref{qfitermtwo}) and Eq.~(\ref{crossterm}) together we can observe that the QFI of $|\Psi_{\chi,\varsigma}\rangle$ is:
\begin{equation} \label{qfi_full}
\begin{split}
\mathcal{F}(|\Psi_{\chi,\varsigma}\rangle)& = 4\left(\langle \partial_\chi{\Psi}_{\chi,\varsigma}|\partial_{\chi}{\Psi}_{\chi,\varsigma}\rangle- |\langle \Psi_{\chi,\varsigma} |\partial_\chi{\Psi}_{\chi,\varsigma}\rangle|^2 \right) \\
& = 4 (\partial_\chi \theta)^2 \left[ \langle \psi,0|U_1^{-1}\hat {H}_2^2 U_1|\psi,0\rangle -  |\langle \psi,0|U_1^{-1}\hat H_2 U_1|\psi,0\rangle|^2 \right] \\
& \hspace{10pt}+ 4 \left[ (\partial_\chi\eta) (\partial_\eta \xi) \right]^2 \langle \psi,0|\hat H^2_1|\psi,0\rangle
\end{split}
\end{equation}
where the first line of second equality is equal to the QFI for phase estimation multiplied by a factor of $(\partial_\chi\theta)^2$ associated with changing variables from the phase $\theta$ to $\chi$:
\begin{equation} \label{phase_qfi}
\begin{split}
\langle \partial_\theta{\Psi}_{\chi,\varsigma}|\partial_{\theta}{\Psi}_{\chi,\varsigma}\rangle- |\langle \Psi_{\chi,\varsigma} |\partial_\theta{\Psi}_{\chi,\varsigma}\rangle|^2 & = \langle \psi,0|U_1^{-1}\hat {H}_2^2 U_1|\psi,0\rangle -  |\langle \psi,0|U_1^{-1}\hat H_2 U_1|\psi,0\rangle|^2 \\
& =  - \langle\hat{n}\rangle_{\text{in}} (1-\eta)\eta(\varsigma -1)^2 + \Var_{\text{in}}({\hat{n}})(\eta + \varsigma - \eta\varsigma)^2
\end{split}
\end{equation}
where the second line has been calculated using the similarity transformations in Eq.~(\ref{phase_loss_similarity}). Similarly, the expression on the second line of the second equality in Eq.~(\ref{qfi_full}) is the QFI for loss estimation multiplied by a factor of $(\partial_\chi\eta)^2$ associated with changing variables from transmissivity $\eta$ to $\chi$:
\begin{equation} \label{loss_term_min}
\begin{split}
\langle \partial_\eta{\Psi}_{\chi,\varsigma}|\partial_{\eta}{\Psi}_{\chi,\varsigma}\rangle- |\langle \Psi_{\chi,\varsigma} |\partial_\eta{\Psi}_{\chi,\varsigma}\rangle|^2 & =   (\partial_\eta \xi)^2 \langle \psi,0|\hat H^2_1|\psi,0\rangle  \\
& = \frac{1}{4\eta(1-\eta)}\langle\hat{n}\rangle_{\text{in}}.
\end{split}
\end{equation}
As this expression is independent of $\varsigma$, the QFI of the purified state $\mathcal{F}(\ket{\Psi_{\chi,\varsigma}})$ is minimised by minimising the term related to the phase information in Eq.~(\ref{phase_qfi}). This can be minimised by differentiating with respect to $\varsigma$ and setting the resultant expression to zero yielding:
$$
\varsigma_\text{opt}  = 1-\frac{\Var_{\text{in}}(\hat{n})}{(1 - \eta)\Var_{\text{in}}(\hat n)+\eta\langle\hat n\rangle_{\text{in}}}
$$
corresponding to the minimised expression:
\begin{equation} \label{phase_term_min}
\frac{\langle\hat n\rangle_{\text{in}}\eta}{(1-\eta) + \langle\hat n\rangle_{\text{in}}\eta/\Var_{\text{in}}(\hat{n})}\leq \frac{\eta\langle\hat n\rangle_{\text{in}}}{1-\eta}.
\end{equation}
Putting together Eqs.~(\ref{qfi_full}), (\ref{loss_term_min}), (\ref{phase_term_min}) we can see the QFI of the state $\varrho_\chi = \Lambda_\chi\left(|\psi\rangle\right)$ after the channel $\Lambda_\chi$ is bounded by:
\begin{equation}
\begin{split}
\mathcal{F}(\varrho_{\chi})&\leq \min_{\varsigma} \left[ \mathcal{F}(\ket{\Psi_{\chi,\varsigma}}) \right] \\
&=(\partial_\chi\theta)^2 \bigg[\frac{4\eta \langle \hat n\rangle_{\text{in}} \Var_{\text{in}}(\hat n)}{(1-\eta)\Var_{\text{in}}(\hat{n})+\eta\langle\hat{n}\rangle_{\text{in}}}\bigg] \\
&\hspace{20pt}+(\partial_\chi\eta)^2\frac{\langle\hat{n}\rangle_{\text{in}}}{\eta(1-\eta)} \\
&\leq\langle\hat{n}\rangle_{\text{in}}\frac{4\eta^2(\partial_\chi\theta)^2+(\partial_\chi\eta)^2}{\eta(1-\eta)}.
\end{split}
\end{equation}

\section{Supplementary Material C: Saturating the quantum limits with finite mean photon number input states} \label{finite_n}
In this section we show that probe states can have a finite mean number of photons and saturate the quantum limits on both CPLE, such that $\mathcal{F}(\varrho) = \mathcal{Q}_\chi$, and DAE, such that $\mathcal{F}(\varrho) = \mathcal{Q}_\eta$.
To demonstrate this mathematical statement, and not in search or practical strategies, we consider a probe state of the form:
\begin{equation}
\begin{split}
\varrho_{p} = \lim_{n \rightarrow \infty} \left[ \frac{p}{n} |\psi_{n}\rangle \langle\psi_{ n}|\otimes|\logic{1}\rangle  \langle \logic{1}| + \left(1-\frac{p}{n}\right)|0\rangle \langle 0|\otimes|\logic{0}\rangle \langle \logic{0}| \right]
\end{split}
\end{equation}
in which the first system is the state of an optical mode incident upon the sample and the second system is a qubit. The state $|\psi_n\rangle$ has a mean number photon number of $n$ and $|0\rangle$ denotes the vacuum. Therefore $\varrho_p$ contains an average of $p$ photons. In the main text of this chapter it was shown that there exist states with a QFI that is asymptotically equivalent to the quantum limit for both CPLE and DAE. We consider $\ket{\psi_n}$ to be a state which is asymptotically equivalent to the quantum limit i.e. $\lim_{n\rightarrow\infty} [\mathcal{F}(\ket{\psi_n})/\mathcal{Q}_\bullet]=1$ where here $\mathcal{Q}_\bullet$ is the quantum limit on $n$ mean photon number states. Due to the linearity of QFI over terms which have support on orthogonal subspaces~\cite{dorner2009}, and the fact that the quantum limit is directly proportional to the mean number of photons in the input state, the QFI of $\varrho_p$ is
$$
\mathcal{F}(\varrho_p)= \lim_{n\rightarrow\infty}\left[(\mathcal{Q}_\bullet/p)n\times (p/n)\right]=\mathcal{Q}_\bullet
$$
where here $\mathcal{Q}_\bullet$ is the quantum limit on $p$ mean photon number input states. We note that $ \ket{\psi_n}$ could be a Gaussian state making this strategy equivalent to the ones discussed in the main text except sometimes no probe photons are used; such a change can only yield a superficial advantage and it will require more trials before the regular Cr\'{a}mer--Rao is saturated; therefore to make a precise measurement this type of modification will not reduce the mean number of photons which are incident upon the sample. The purpose of this section is merely to show that the bounds on QFI which we have found are tight.

\section{Supplementary Material D: Optimising $\mathcal{D}$}
In this appendix we optimize $\mathcal{D} = \|\partial_\chi\tilde{\bm{d}}\|^2\big/\tilde{V}_{\text{min}}$, for a state which is squeezed in the optimal direction, given in the main text Eq.~(8) 
over the amount of squeezing $n_\text{sq}$ (or equivalently $r$) for a given mean number of photons $\langle\hat n_0 \rangle$. Restating $\mathcal{D}$:
\begin{equation}
\mathcal{D}=(\langle\hat{n}\rangle_\text{in}-n_{\text{sq}})\frac{4\mathcal{S}_\chi^2}{\left[\e^{-2r}\eta+(1-\eta)\right]}.
\end{equation}
we note that $\mathcal{D}$ is will be optimized by the same value of $n_\text{sq}$ whether the speed of the transfer amplitude is due its phase dependence or loss dependence. In Ref.~\cite{birchall2017quantum} $\|\partial_\theta\tilde{\bm{d}}\|^2\big/\tilde{V}_{\text{min}}$ was optimised over $n_\text{sq}$ for a state squeezed in the optimal direction. As the additional loss dependence of the displacement vector does not change the optimal amount of squeezing, the optimised value of:
\begin{equation}
\begin{split}
n_{\text{sq}}=& \frac{\left(\sqrt{1-4 (\eta -1) \eta  \langle\hat{n}\rangle_{\text{in}}}-1\right)^2}{4 (1-\eta) \left(\sqrt{1-4 (\eta -1) \eta  \langle\hat{n}\rangle_{\text{in}}}-\eta \right)} \\
=& \mathcal{O}\left(\sqrt{\langle\hat{n}\rangle_{\text{in}}}\right)
\end{split}
\end{equation}
for phase estimation reported in~\cite{birchall2017quantum} will be the optimal value for CPLE. Inserting this value for $n_\text{sq}$  into the main text Eq.~(8) 
 yields
\begin{equation}
\mathcal{D} =\mathcal{Q}_\chi\,\frac{2 (\eta -1) \langle\hat{n}\rangle_{\text{in}}+\sqrt{1-4 (\eta -1) \eta  \langle\hat{n}\rangle_{\text{in}}}-1}{2 (\eta -1) \langle\hat{n}\rangle_{\text{in}}}.
\end{equation}
This expression was used to generate Figure~2~a). 

\section{Supplementary Material E: Optimising $\mathcal{N}$} \label{sec:opt_n}
In this section we optimise $\mathcal{N}$ for an amplitude squeezed Gaussian with a finite mean number of photons. Here we restate $\mathcal{N}$:
\begin{equation}
\begin{split}
1/\Delta^2\eta&=\left(\partial_\eta\langle\hat{n}\rangle_{\text{out}}\right)^2\big/\Var_{\text{out}}({\hat{n}})\equiv  \mathcal{N}\\
&=(\langle\hat{n}\rangle_{\text{in}})^2\big/ \left[\eta^2\Var_{\text{in}}({\hat{n}})+\eta(1-\eta)\langle\hat{n}\rangle_{\text{in}}\right].
\end{split}
\end{equation}
As remarked in the main paper when discussing direct absorption estimation (DAE), the task of minimising $\mathcal{N}$ for a given $\langle\hat{n}\rangle_{\text{in}}$ is equivalent to minimising the number variance of the input state. Using the expressions given in \cite{dodonov1994} we arrive at the number variance of an amplitude squeezed pure Gaussian state:
\begin{equation}
\Var_{\text{in}}({\hat{n}}) = 2 \langle\hat{n}\rangle_{\text{in}}  n_\text{sq}-2 \langle\hat{n}\rangle_{\text{in}} \sqrt{ n_\text{sq} ( n_\text{sq}+1)}+\langle\hat{n}\rangle_{\text{in}}+2 \sqrt{ n_\text{sq}^3
	( n_\text{sq}+1)}+ n_\text{sq}.
\end{equation}
The above expression can be minimised, for a given $\langle\hat{n}\rangle_{\text{in}}$, by differentiating with respect to $n_\text{sq}$ and equating to zero. The resulting equation cannot be solved analytically for $n_\text{sq}$ but it can be solved for $\langle\hat{n}\rangle_{\text{in}}$:
\begin{equation}
\langle\hat{n}\rangle_{\text{in}} = \left(2  n_\text{sq}+2 \sqrt{ n_\text{sq} ( n_\text{sq}+1)}+1\right) \left( n_\text{sq} (4
 n_\text{sq}+3)+\sqrt{ n_\text{sq} ( n_\text{sq}+1)}\right).
\end{equation}
This expression is clearly monotonic in $n_\text{sq}$ for $n_\text{sq}\geq 0$, therefore for a given $\langle\hat{n}\rangle_{\text{in}}$ the optimal value of $n_\text{sq}$ will be unique and can easily be found using numerical methods. This was the approach used to generate the curves in Figure~2~b). 

\section{Supplementary Material F: Multi-pass Strategies} \label{multipass_phase_and_loss}

In this section of the appendix we consider the utility of multi-pass interrogation techniques for CPLE and DAE using techniques analogous to those presented in the main paper. First we will consider CPLE. The application of a channel, which imparts a phase $\theta(\chi)$ has a transmissivity of $\eta$, $k$ times in series results in an overall transition amplitude of $T(k) = (\sqrt{\eta}\e^{\ti \theta})^k$. The square of the speed of this transition amplitude is
\begin{equation}
\begin{split}
\mathcal{S}_\chi^2(k)& = |\partial_\chi T(k)|^2 \\
& = \eta^{k-1} k^2 [\mathcal{S}_\chi(k=1)]^2
\end{split}
\end{equation}
where $\mathcal{S}_\chi(k=1)$ is the speed of the transition amplitude for a single application of channel.  We can insert this modified transition amplitude speed, together with a modified transmissivity of $\eta^k$, into Eq.~(7) 
 of the main text to find the capabilities of classical states and into the final line of Eq.~(4) 
 of the main text to find the capabilities of optimal quantum states for CPLE in multi-pass set-ups. Since the capabilities of classical states and the capabilities of optimal quantum states are modified by the same factor due to the multiple passes, independently of whether it is phase or loss variation contributing to the speed of the transition amplitude, we can conclude that multi-pass strategies will be just as effective for CPLE as for lossy phase estimation in the preceding chapter. We can also conclude that the optimal number of passes will be the same for CPLE as they were for lossy phase estimation. Specifically, when there is no loss introduced by the apparatus, only by the sample itself, the number of passes used to maximise the precision for a given number of lost photons should be adjusted such that $\eta^{k_\text{opt}} \approx 20\%$ (the analytic expression is given in Birchall \textit{et al.}~\cite{birchall2017quantum}) in order to maximise the precision per lost photon. The ratio of achievable precision between optimal quantum multi-pass strategies and classical multi-pass strategies for CPLE with a given number of incident photons will also be the same as it is for lossy-phase estimation. Therefore the maximum reduction in RMSE one can achieve for CPLE using optimal quantum techniques is $\approx 20\%$~\cite{birchall2017quantum}.

If the elements used to do perform a multi-pass strategy have additional inefficiencies encountered during state preparation $\eta_p$, state detection $\eta_d$ and in each round trip $\eta_r$, then the speed of the overall transition amplitude will be reduced by a factor of $\sqrt{\eta_p \eta_d \eta_r^{k-1}}$. As before, since this factor is independent of whether the initial speed of the transition amplitude is due to phase variation or loss variation, CPLE is affected by these additional losses in the same way that lossy phase estimation is. Therefore we can conclude that for CPLE with imperfect components, as long as the round trip loss $1-\eta_r$ is less than the combined state preparation and detection loss $\eta_r>\eta_p\eta_d$, then non-classical techniques cannot provide more than a $20\%$ reduction in RMSE over classical techniques. This was shown for lossy phase estimation in~\cite{birchall2017quantum}.

Multi-pass strategies also enhance DAE in the same way that CPLE is enhanced. The limit on QFI for DAE can be obtained from the limit on QFI for CPLE by setting $(\partial_\chi \theta = 0)$ and $(\partial_\chi \eta = 1)$ such that $\mathcal{Q}_\eta = \mathcal{Q}\displaystyle{\big|_{\partial_\chi\theta=0,\,\partial_\chi\eta=1}}$. Similarly, the capabilities of quantum states in combination with multi-pass setups for DAE are given by the the capabilities of quantum states in combination with multi-pass setups for CPLE when $(\partial_\chi \theta = 0)$ and $(\partial_\chi \eta = 1)$. The capabilities of classical states and multi-pass setups is also given by the expressions for CPLE when $(\partial_\chi \theta = 0)$ and $(\partial_\chi \eta = 1)$. We deduce that the same conclusions which were drawn about the utility of multi-pass setups for CPLE also apply to DAE. For example, the optimal number of passes through the sample will be the same for DAE as it is for CPLE, the precision enhancements allowed by non-classical strategies will be the same for DAE as it is for CPLE and the inclusion of imperfect components will also have the same effect on multi-pass DAE strategies as it does for multi-pass CPLE strategies.

\end{document}